\documentclass[twocolumn,aps,showpacs,superscriptaddress,prl]{revtex4}

\usepackage{amsmath,amssymb,amsmath}
\usepackage{graphicx}
\usepackage{dcolumn}
\usepackage{bm}

\begin{document}

\title{Exploiting the composite character of Rydberg atoms for cold atom trapping}

\author{Michael Mayle}
\affiliation{Theoretische Chemie, Universit\"at Heidelberg, D-69120 Heidelberg, 
Germany}

\author{Igor Lesanovsky}
\affiliation{Institut f\"ur Theoretische Physik,
Universit\"at Innsbruck, A-6020 Innsbruck, Austria}

\author{Peter Schmelcher}
\affiliation{Theoretische Chemie, Universit\"at Heidelberg, D-69120 Heidelberg, 
Germany}
\affiliation{Physikalisches Institut, Universit\"at Heidelberg,
 D-69120 Heidelberg, Germany}

\date{\today}

\begin{abstract}
By investigating the quantum properties of magnetically trapped 
$nS_{1/2}$ Rydberg atoms, it is demonstrated that the composite nature 
of Rydberg atoms significantly alters their trapping properties
opposed to point-like particles with the same magnetic moment.
We show how the specific signatures of the Rydberg trapping potential
can be probed by means of ground state atoms that are off-resonantly 
coupled to the Rydberg state via a two photon laser transition.
In addition, it is demonstrated how this approach provides new 
possibilities for generating traps for ground state atoms.
Simulated time-of-flight pictures mirroring the experimental 
situation are provided.
\end{abstract}

\pacs{32.10.Ee, 
32.80.Ee, 
32.60.+i, 
37.10.Gh  
}

\maketitle

The size of Rydberg atoms can easily exceed that of ground state atoms 
by several orders of magnitude. Already a state with principal quantum 
number $n\approx 40$ has an electronic orbit that measures $\sim 200$ 
nm in diameter and thus is more than thousand times larger than the 
ground state \cite{gallagher}. 
The associated displacement of the atomic charges makes Rydberg atoms
highly susceptible to external fields and at the same time is the 
origin of their strong mutual interaction. 
The latter has been demonstrated to entail a blockade mechanism
\cite{blockade}
thereby effectuating collective Rydberg excitations in ultracold gases
\cite{rydexp}
and facilitating the 
observation of conditional dynamics between two single atoms separated 
by a few $\mu$m \cite{singlerydexp}.

Owed to their large size, Rydberg atoms do not only interact much 
stronger than their ground state counterparts but also behave quite 
differently when placed in electric and/or magnetic field configurations
that provide trapping for ground state atoms.
Several works have focused on this issue, discussing traps for Rydberg 
atoms based on electric \cite{hyafil:103001}, optical \cite{dutta00}, 
or strong magnetic fields \cite{choi:243001}. 
As opposed to ground state atoms, the electronic dynamics usually does 
not decouple from the external center of mass (c.m.) motion and trapping
becomes a particular delicate task. 
This becomes evident in tight magnetic traps, where the point particle 
description of the atoms breaks down \cite{hezel,mayle:113004} as the 
extension of the Rydberg state becomes comparable or even larger than 
the length scale imposed by the trap, i.e., the extension of the c.m.\ 
wave function.

In this letter we illuminate the question of how the composite character
of Rydberg atoms, i.e., the fact that it consists of an outer electron 
far away from a compact ionic core, becomes manifest in standard magnetic
traps. 
In particular, we focus on $nS_{1/2}$ Rydberg states which are excited 
from an atomic gas of $^{87}$Rb atoms being in the $5S_{1/2},F=m_F=2$ 
ground state, as done experimentally in 
Refs.~\cite{rydexp}.
We demonstrate that, although both states possess the same magnetic 
moment, the trapping potential of the Rydberg states is substantially 
altered due the composite nature of the atom.
This effect can be 
directly measured by off-resonantly coupling the ground and the 
Rydberg state via excitation lasers. The subsequent evolution of the
dressed ground state atoms is qualitatively modified by the admixture 
of the Rydberg potential surface leaving its imprint in the 
time-of-flight (TOF) expansion image. In this context we show that 
dressing ground state atoms with Rydberg states has the potential to 
become a useful method to design trapping potentials which are not 
simply achievable by means of magnetic fields.

As the basic ingredient for magnetically trapping Rydberg atoms,
we consider the Ioffe-Pritchard (IP) field configuration given by
$\mathbf{B}(\mathbf{x})=\mathbf B_c+\mathbf{B}_l(\mathbf{x})$
with
$\mathbf B_c=B\mathbf e_3$, $\mathbf{B}_l(\mathbf{x})=
G\left[x_1\mathbf{e}_1-x_2\mathbf{e}_2\right]$.
The corresponding vector potential reads
$\mathbf{A}(\mathbf{x})=
\mathbf{A}_c(\mathbf{x})+\mathbf{A}_l(\mathbf{x})$,
with
$\mathbf{A}_c(\mathbf{x})=
\frac{B}{2}\left[x_1\mathbf{e}_2-x_2\mathbf{e}_1\right]$
and $\mathbf{A}_l(\mathbf{x})=Gx_1x_2\mathbf{e}_3$;
$B$ and $G$ are the Ioffe field strength and the gradient, respectively.
Along the lines of Ref.~\cite{hezel},
we model the mutual interaction of the highly excited valence electron 
and the remaining closed-shell ionic core of a Rydberg atom by an 
effective potential which depends only on the distance of the two 
particles.
After introducing relative and c.m.\ coordinates ($\mathbf{r}$ and
$\mathbf{R}$) and employing the unitary transformation
$U=\exp\left[\frac{i}{2}(\mathbf{B}_c\times \mathbf{r}) \cdot
\mathbf{R}\right]$, the Hamiltonian describing the Rydberg atom becomes
(atomic units are used unless stated otherwise)
\begin{eqnarray}
\label{eq:hamfinaluni}
H&=&H_A+\frac{\mathbf{P}^2}{2M}
+\tfrac{1}{2}[\mathbf L_r+2\mathbf S]\cdot\mathbf B_c
+\mathbf S\cdot\mathbf{B}_l(\mathbf{R+r})
\nonumber\\
&&+\mathbf{A}_l(\mathbf{R+r})\cdot\mathbf{p}
+\tfrac{1}{2}\mathbf{A}_c(\mathbf{r})^2+H_\mathrm{corr}\,.
\end{eqnarray}
Here, $H_A=\mathbf{p}^2/2+V_l(r)+V_{so}(\mathbf{L}_r,\mathbf{S})$
is the field-free Hamiltonian of the valence electron 
whose core penetration, scattering, and polarization effects
are accounted for by the $l$-dependent model potential $V_l(r)$
(see Ref.~\cite{PhysRevA.49.982} for its explicit form)
while $\mathbf L_r$ and $\mathbf S$ denote its orbital 
angular momentum and spin, respectively.
The spin-orbit interaction is given by
$V_{so}(\mathbf{L}_r,\mathbf{S})=
\frac{\alpha^2}{2}\big(1-\tfrac{\alpha^2}{2}V_l(r)\big)^{-2}
\frac{1}{r}\frac{\mathrm d V_l(r)}{\mathrm d r}
\mathbf L_r\cdot\mathbf S$; 
the term $(1-\alpha^2V_l(r)/2)^{-2}$
has been introduced to regularize the nonphysical divergency near the 
origin \cite{condon35}.
$H_\mathrm{corr}=-\boldsymbol{\mu}_c\cdot \mathbf{B(R)}+
\frac{1}{2}\mathbf A_l(\mathbf{R+r})^2
+\frac{1}{M}\mathbf B_c\cdot(\mathbf{r\times P})
+U^\dagger[V_l(r)+V_{so}(\mathbf{L}_r,\mathbf{S})]U$
are small corrections which can be safely neglected in the 
parameter regime we are focusing on; $\mbox{\boldmath$\mu$}_c$
is the magnetic moment of the nucleus.
Note that, in contrast to the low angular momentum states discussed
in this letter, mostly circular states are
considered in Refs.~\cite{hezel,mayle:113004} 
which allows employing a purely Coulombic potential and
neglecting the fine structure.
Since the $Z$-component of the c.m.\ momentum commutes with the
Hamiltonian (\ref{eq:hamfinaluni}), the longitudinal motion can be
integrated out by employing plane waves.
In order to solve the remaining coupled Schr\"odinger equation, we 
employ a Born-Oppenheimer separation of the c.m.\ motion and the 
electronic degrees of freedom.
We are thereby led to an electronic Hamiltonian for fixed c.m.\ position
of the atom whose eigenvalues $E_\kappa(\mathbf R)$ depend 
parametrically on the c.m.\ coordinates. These adiabatic electronic 
surfaces serve as potentials for the quantized c.m.\ motion.
The emerging non-adiabatic (off-diagonal) coupling terms 
between different electronic states can be neglected in our parameter 
regime \cite{hezel}.

For fixed total electronic angular momentum $\mathbf{J=L}_r+\mathbf S$,
approximate expressions for the adiabatic electronic energy surfaces 
can be derived. For this reason, we reformulate the electronic 
Hamiltonian as \cite{hezel}
\begin{eqnarray}
\label{eq:helec}
H_e&=&H_A+\frac{1}{2}\left[\mathbf{L}_r+2\mathbf{S}\right]\cdot
\mathbf B(\mathbf R)+GXYp_z+H_r
\end{eqnarray}
where $H_r=\mathbf{A}_l(\mathbf{r})\cdot\mathbf{p}+
\mathbf{B}_l(\mathbf{r})\cdot\mathbf{S}+
\tfrac{1}{2}\mathbf{A}_c(\mathbf{r})^2$
only depends on the relative coordinate. 
Since in the desired parameter regime $H_r$ solely represents
an energy offset to the electronic energy
surfaces, we will omit its contribution in the following.
The first two terms of the Hamiltonian~(\ref{eq:helec}) can be 
diagonalized analytically by applying the spatially dependent 
transformation $U_r=e^{-i \gamma (L_x+S_x)}e^{-i \beta (L_y+S_y)}$
which rotates the $z$-axis into the local magnetic field direction.
The corresponding rotation angles are defined by
$\sin\gamma =-GY/\sqrt{B^2+G^2(X^2+Y^2)}$ and
$\sin\beta =-GX/\sqrt{B^2+G^2X^2}$.
The transformed Hamiltonian then becomes
\begin{eqnarray}
 U_r H_e U_r^\dagger =H_A+\tfrac{1}{2}g_jJ_z\sqrt{B^2+G^2(X^2+Y^2)}
+H'
\label{eq:htrans}
\end{eqnarray}
with $g_j=\frac{3}{2}+\frac{s(s+1)-l(l+1)}{2j(j+1)}$ and
$H'=GXYU_r p_z  U_r^\dagger$.
Like for ground state atoms, the second term of Eq.~(\ref{eq:htrans})
represent the coupling of a point-like particle to the magnetic field
via its magnetic moment $\boldsymbol{\mu}=\frac{1}{2}\mathbf L_r
+\mathbf S$.
For a given state $|\kappa\rangle=|nljm_j\rangle$
with field-free electronic energy $E_\kappa^{el}$,
this gives rise to the electronic 
potential energy surface 
$E_\kappa^{(0)}(\mathbf R) =E_\kappa^{el}+\tfrac{1}{2}g_jm_j|\mathbf{B(R)}|$,
which is rotationally symmetric around the $Z$-axis and confining for 
$m_j>0$.
For small radii ($\rho=\sqrt{X^2+Y^2}\ll B/G$) an expansion up to second 
order yields a harmonic potential
$E_\kappa^{(0)}(\rho)\propto \tfrac{1}{2}M\omega^2\rho^2$
with a trap frequency of $\omega=G\sqrt{g_jm_j/2MB}$.
At this point, let us briefly comment on the role of the hyperfine 
interaction. For Rydberg atoms and the regime of field strengths we are
considering, the hyperfine interaction can be treated perturbatively
while for ground state atoms, on the other hand, 
one must consider the total angular momentum $\mathbf{F=J+I}$
($\mathbf I$ being the nuclear spin) yielding energy surfaces 
$\sim  \frac{1}{2}g_Fm_F|\mathbf{B(R)}|$ with
$g_F=g_j\frac{F(F+1)+j(j+1)-I(I+1)}{2F(F+1)}$.
As a consequence, the $nS_{1/2}, m_j=1/2$ Rydberg and $5S_{1/2}$
ground state energy surface are identical for $F=m_F=2$, which we
consider in the following.

\begin{figure}
\hspace{-0.5cm}
 \includegraphics[scale=1.0]{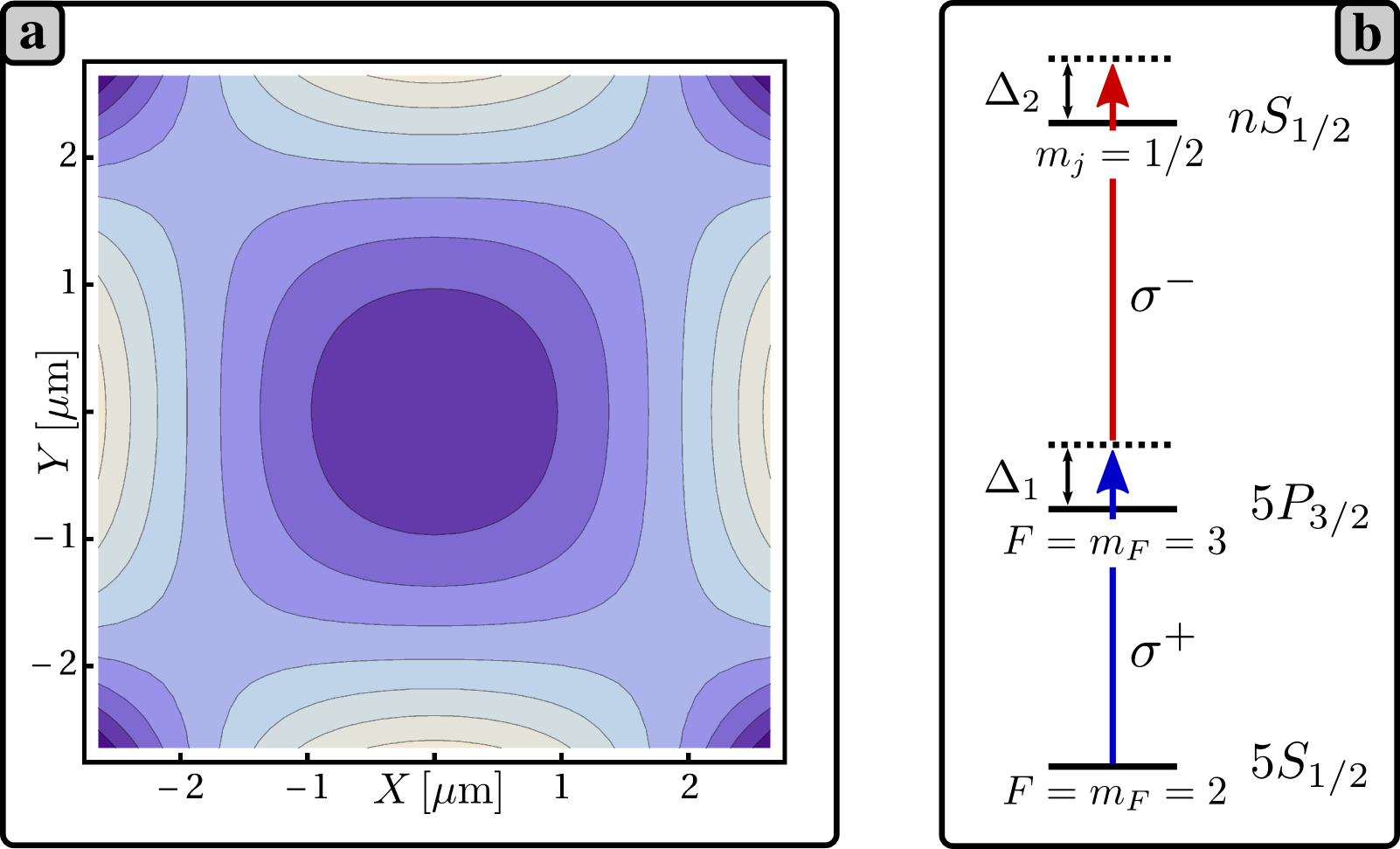}
\caption{
(a) Contour plot of the $40S_{1/2}, m_j=1/2$ Rydberg surface for $B=1$ G
and $G=10$ Tm$^{-1}$. The loss of the azimuthal symmetry introduced by 
the composite nature of the Rydberg atom as given by Eq.~(\ref{eq:gxy}) 
is evident.
The depth of the minimum along the diagonals corresponds 
to $\sim 18\,\hbar\omega$, $\omega=2\pi\times1.3$ kHz being 
the trap frequency of the harmonic confinement at the origin.
(b) Idealized level scheme for an off-resonant two photon coupling of 
the ground and Rydberg state of $^{87}$Rb. In a IP trap, additional 
atomic levels and polarizations contribute away from the trap center, 
see text.
}
\label{fig1}
\end{figure}

Within the given approximations, the composite character of the Rydberg 
atom is manifest in the remaining term $H'$ of the Hamiltonian 
(\ref{eq:htrans}).
It contributes only for large radii close to the diagonal $X=Y$
and can be treated perturbatively. 
While it vanishes in first order, second order
perturbation theory yields
\footnote{We approximated $p_z=i[H_A,z]-i[V_l(r)+V_{so},z]
\approx i[H_A,z]$ since the second term is energetically suppressed 
in perturbation theory.}
\begin{eqnarray}
 E_\kappa^{(2)}(\mathbf R)
= G^2X^2Y^2\sum_{\kappa'\neq\kappa}
(E_\kappa^{el}-E_{\kappa'}^{el})\times\big|\langle \kappa|U_rzU_r^\dagger
|\kappa'\rangle\big|^2.
\label{eq:gxy}
\end{eqnarray}
Since $E_\kappa^{(0)}(\mathbf R)$ resembles the confinement of
ground state atoms, we will attribute
$E_\kappa^{(2)}(\mathbf R)$ to the composite nature of the Rydberg
atom.
For $nS_{1/2}$ Rydberg states, Eq.~(\ref{eq:gxy}) gives rise to
$E_\kappa^{(2)}(\mathbf R)=-C\cdot G^2X^2Y^2$ 
where $C=0.48$ for $n=40$.
One particular property of the additional contribution 
$E_\kappa^{(2)}(\mathbf R)$ is its spatial dependence:
Unlike $E_\kappa^{(0)}(\mathbf R)$ it does not preserve the
azimuthal symmetry. 
In Fig.~\ref{fig1}(a) this behaviour is illustrated for
the example of $n=40$, $B=1$ G, $G=10$ Tm$^{-1}$.
Moreover, far from the trap center, but close to the diagonal $X=Y$,
the character of the potential surface
eventually changes from trapping to anti-trapping.
Hence, by effectively altering the trap geometry the composite 
nature distinguishes the Rydberg atom from its point-like 
counterparts.
We remark that the relative difference between the 
analytical energy surface $E_\kappa^{(0)}(\mathbf R)+E_\kappa^{(2)}(\mathbf R)$
and the numerical diagonalization of Hamiltonian (\ref{eq:hamfinaluni})
is less than $1\%$ for the case of Fig.~\ref{fig1}(a);
for smaller $G$ or higher $B$ the agreement is even better.
Hence, the validity of the perturbative approach is ensured for
a wide range of field strengths.

Because of the finite lifetime of Rydberg atoms and their strong
susceptibility to external perturbations (stray electric fields, 
mutual interactions etc.), the experimental observation of the 
composite character might be a difficult task.
However, these restrictions can be alleviated by employing the following
scheme.
Instead of completely transfering a ground state $^{87}$Rb atom to 
a certain $nS_{1/2}$ Rydberg state, we rather couple it off-resonantly 
by a two photon laser transition.
This procedure results in a dressed ground state atom whose trapping
potential is effectively altered by the admixture of the Rydberg surface.
Any change in the ground state c.m.\ wave function can then be attributed 
to the composite nature of the Rydberg atom since 
both possess the same magnetic moment.

Let us investigate the excitation scheme that is frequently encountered 
in experiments \cite{rydexp}:
Laser 1, which is $\sigma^+$ polarized, drives the transition 
$s\rightarrow p$ detuned by $\Delta_1$ while a second, $\sigma^-$ 
polarized laser then couples to the Rydberg state 
$n\equiv nS_{1/2},F=m_F=2$, with $s$ denoting the ground state 
$5S_{1/2}, F=m_F=2$ and $p$ the intermediate state $5P_{3/2},F=m_F=3$.
The complete two photon transition is supposed to be off-resonant
by $\Delta_2$, c.f., Fig.~\ref{fig1}(b).
In the dipole approximation, the interaction of an atom with a laser 
field is given by 
$H_l=\boldsymbol{\epsilon}\cdot \mathbf r\,E_0\cos\omega t$
with $\boldsymbol{\epsilon}$ being the polarization vector of the 
excitation laser.
However, since in a IP trap the quantization axis is spatially 
dependent the polarization vector $\boldsymbol{\epsilon}$ is only 
well-defined as $\sigma^+$ or $\sigma^-$ at the trap center.
In the rotated frame of reference, i.e., after
applying the unitary transformation $U_rH_lU_r^\dagger$,
contributions of all polarizations emerge.
Hence, the excitation scheme becomes more involved: Intermediate states
with $F,m_f\in[1,3]$ contribute as well, while on the Rydberg side also
$m_j=-1/2$ becomes accessible.
Altogether, this yields an excitation scheme including 12 atomic levels.
In the following, we distinguish two complementary regimes:
(a) strong Ioffe fields where simplifications lead to an analytical 
solution, and (b) high gradients where a numerical treatment is 
indispensable.
In both cases, the time-dependence of the Hamiltonian is removed by 
employing the rotating wave approximation while the intermediate levels 
are eliminated by a strong off-resonance condition.

\begin{figure}
 \includegraphics{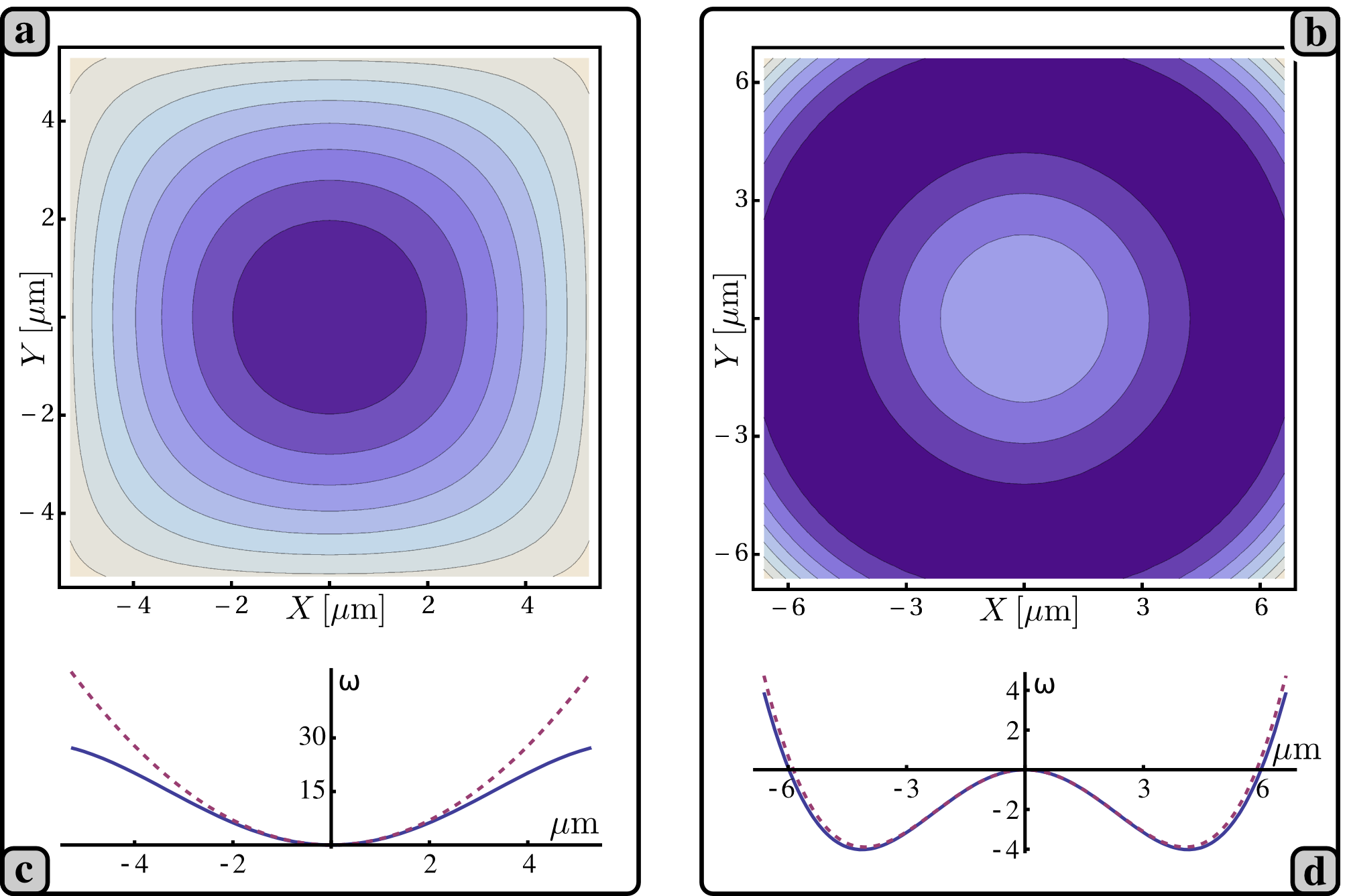}
 \caption{
(a) Contour plot of the dressed ground state surface $E_-$ for 
$n=40$, $B=10$ G, and $G=5$ Tm$^{-1}$. 
(c) Section along $X=Y$ of the same energy surface.
The deviation from the harmonic confinement
of non-dressed ground state atoms (dashed line)
due to the composite nature of the Rydberg atom is evident.
The energy scale is given in terms of the trap frequency 
$\omega=2\pi\times200$ Hz.
(b) and (d) The same as on the left panel but for $B=1$ G, $G=5$ 
Tm$^{-1}$.
In this high gradient regime the Rabi frequency and therefore the light 
shift $V_s$ exhibit a strong spatial dependence resulting in the maximum 
at the trap center;
for the dashed line the contribution $E_\kappa^{(2)}(\mathbf R)$ of the 
composite character is omitted.
Without dressing, a ground state atom would experience a trap frequency
of $\omega=2\pi\times638$ Hz.
For values of the Rabi frequencies and detunings see text.}
 \label{fig2}
\end{figure}

Most insights in the underlying physics can be gained in the 
regime of strong Ioffe fields $B$ (and/or weak gradients $G$)
where the quantization axis is predominantly defined by the Ioffe field.
Selection rules then allow us to revert to the more simple 3-level 
scheme $s\leftrightarrow p\leftrightarrow n$ and incorporate the 
residual, albeit rather weak spatial dependence in the one photon Rabi
frequencies
\begin{eqnarray}
 \omega_{ps}&=&\tfrac{1}{2}(\cos\gamma+\cos\beta
-i\sin\gamma\sin\beta)\cdot\omega_{ps}^{(0)}
=\omega_{sp}^*\\
\omega_{np}&=&\tfrac{1}{2}(\cos\gamma+\cos\beta
+i\sin\gamma\sin\beta)\cdot\omega_{np}^{(0)}
=\omega_{pn}^*,
\end{eqnarray}
$\omega_{ij}^{(0)}=E_{0,ij}\langle i|\boldsymbol{\epsilon}_{ij}
\cdot\mathbf r|j\rangle$
being the Rabi frequency at the trap center.
Employing the rotating wave approximation and adiabatically 
eliminating the intermediate $5P_{3/2}$ state eventually 
provides us a 2-level system
\begin{equation}
 \mathcal H_{2l}=
\begin{pmatrix}
 \Delta_2+\tilde{E}_n+V_n& -\Omega/2\\
 -\Omega^*/2&\tilde{E}_s+V_s
\end{pmatrix}
\label{eq:h2l}
\end{equation}
with an effective Rabi frequency of
\begin{eqnarray}
\Omega&=&\frac{\omega_{ps}\omega_{np}}{4}
\left[\frac{1}{\tilde{E}_s-\tilde{E}_p-\Delta_1}
+\frac{1}{\tilde{E}_n-\tilde{E}_p+\Delta_2-\Delta_1}\right]
\end{eqnarray}
and light shifts of
$V_n=-|\omega_{np}|^2/4(\tilde{E}_p-\tilde{E}_n+\Delta_1-\Delta_2)$
and
$V_s=-|\omega_{ps}|^2/4(\tilde{E}_p-\tilde{E}_s+\Delta_1)$.
The involved energy surfaces are given by 
$\tilde{E}_p=2|\mathbf{B(R)}|$, 
$\tilde{E}_s=\frac{1}{2}|\mathbf{B(R)}|$, 
and $\tilde{E}_n=\tilde{E}_s-C\cdot G^2X^2Y^2$, respectively.
Neglecting the composite character 
in the limit $\Delta_1\gg\Delta_2$, one recovers
$\Omega=-\omega_{ps}\omega_{np}/2\Delta_1$, $V_n=
-|\omega_{np}|^2/4\Delta_1$, and
$V_s=-|\omega_{ps}|^2/4\Delta_1$.
The diagonalization of the Hamiltonian (\ref{eq:h2l}) yields the
dressed Rydberg and ground state surfaces $E_\pm$.
For large detunings $\Delta_2\gg\Omega$ one can approximate
\begin{equation}
 E_-\approx
\tilde{E}_s+V_s
-\frac{\Omega^2}{4\Delta_2} 
+\frac{\Omega^2}{4\Delta_2^2}(\tilde{E}_n+V_n-\tilde{E}_s-V_s)\,,
\label{eq:eminus}
\end{equation}
i.e., the contribution of the Rydberg surface to the dressed ground 
state trapping potential is suppressed  by the factor 
$(\Omega/\Delta_2)^2$.
Furthermore, any spatial variation in the light shift $V_s$ and in
the Rabi frequency $\Omega$ will effectively alter the trapping
potential experienced by the dressed ground state atom.
As an example, let us investigate the configuration
$B=10$ G, $G=5$ Tm$^{-1}$, 
$\omega_{ps}^{(0)}=\omega_{np}^{(0)}=2\pi\times30$ MHz,
$\Delta_1=-2\pi\times220$ MHz, and $\Delta_2=-2\pi\times10$ MHz.
In this case, the contribution Eq.~(\ref{eq:gxy}), which derives from 
the composite nature of the Rydberg atom, prevents the Rydberg potential
from supporting any confined c.m.\ state.
This strong deviation from the point-like behaviour 
is consequently mirrored in the dressed ground state potential: 
Along $X=Y$, where the effect of Eq.~(\ref{eq:gxy}) is most pronounced,
the trapping potential is gradually lowered compared to the harmonic
confinement of the non-dressed ground state, c.f., Fig.~\ref{fig2}(c).
As a consequence, the two-dimensonial trapping potential loses its
azimuthal symmetry, see Fig.~\ref{fig2}(a).

\begin{figure}
 \includegraphics[width=8.5cm]{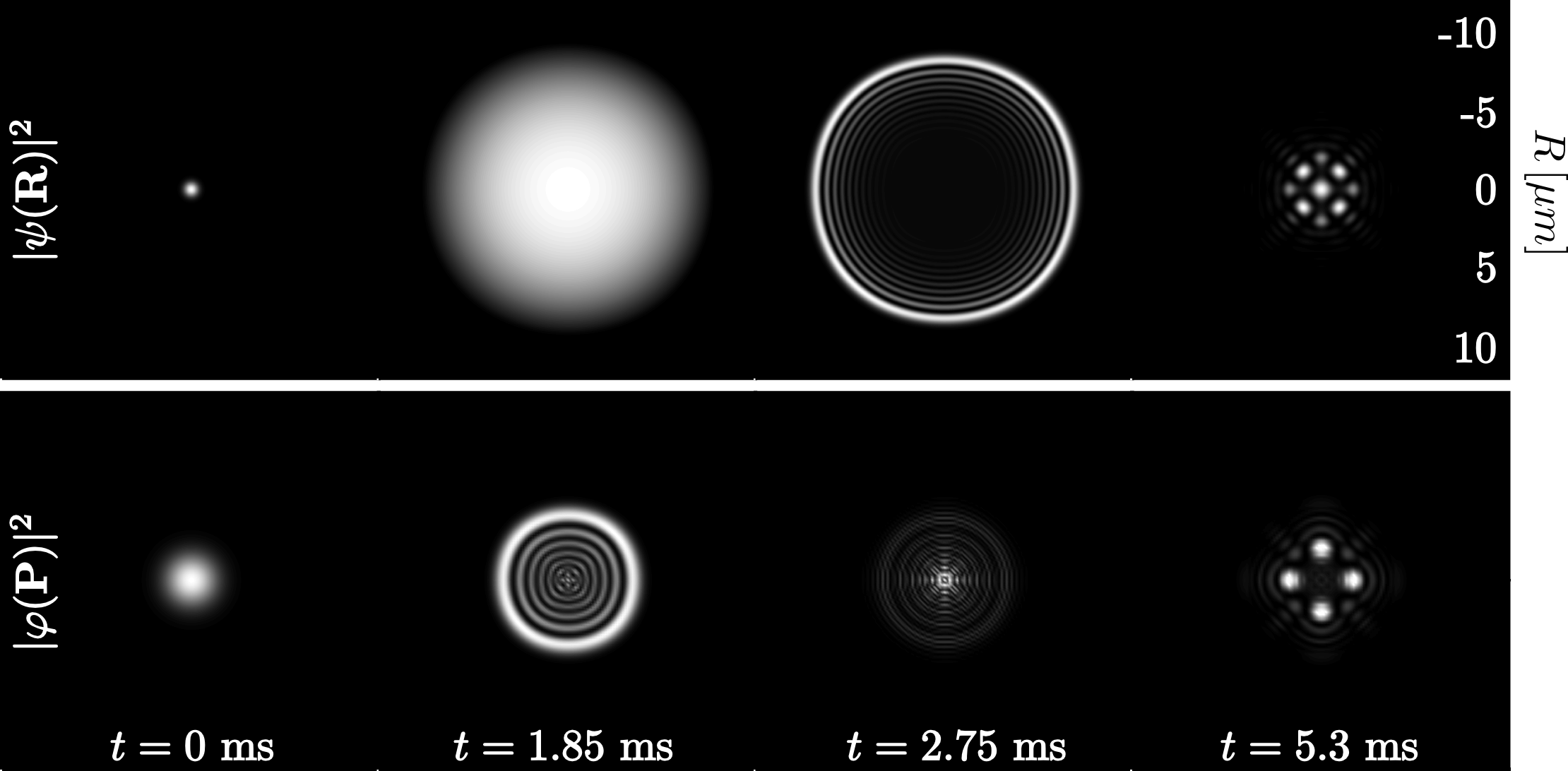}
 \caption{Temporal evolution of the probability density of a dressed 
ground state atom in position (upper row) and momentum (lower row) 
space, respectively.
The momentum distribution simulates TOF expansion measurements.
Being initially in the lowest non-dressed c.m.\ eigenstate
with a trap frequency of $\omega=2\pi\times638$ Hz,
the atom subsequently evolves in the potential given in 
Fig.~\ref{fig2}(b).
The time scale of 5 ms corresponds to the first oscillation of
the c.m.\ wave packet. The loss of the azmimuthal symmetry
due to the composite character of Rydberg atoms is
evident once the wave function has probed the outer parts
of the potential surface ($t>2.75$ ms).}
 \label{fig3}
\end{figure}

Another situation arises for strong gradients where 
the approximation of a 3-level system breaks down
and the full 12-level system must be solved for 
accurate results. 
In Figure \ref{fig2}(b) and (d) the results of such a calculation are 
shown for the specific example of $B=1$ G, $G=5$ Tm$^{-1}$, 
$\omega_{ps}^{(0)}=2\pi\times75$ MHz, 
$\omega_{np}^{(0)}=2\pi\times30$ MHz,
$\Delta_1=-2\pi\times430$ MHz, and $\Delta_2=-2\pi\times50$ MHz.
For this configuration, the light shift $V_s$ develops 
a strong spatial dependence 
which results in a qualitative change of the dressed ground state
potential, namely, from a parabolic to a ring-shaped behaviour.
A wave packet initially prepared in the non-dressed c.m.\ ground state 
disperses in such a potential while being reflected at the outer 
repulsive walls, leading to breathing oscillations.
In this way, the outer parts of the dressed potential surface,
where the composite character of the Rydberg atom becomes manifest, 
are probed.
Simulated TOF expansion images reveal the influence of the Rydberg 
admixture, c.f., Fig.~\ref{fig3}: The observed four-fold symmetry 
after one oscillation ($t=5.3$ ms) is solely due to the contribution 
$E_\kappa^{(2)}(\mathbf R)$; its absence would conserve the azimuthal
symmetry, i.e., the ring shaped pattern.

Let us finally comment on the experimental feasibility.
The proposed dressed states possess a finite effective lifetime
which can be estimated by $\tau=\tau_n/|c|^{2}$ with
$|c|^2=(\Omega/2\Delta_2)^{2}$ being the two-level
admixture coefficient of the Rydberg state and
$\tau_n$ its radiative lifetime.
With $\tau_{40}\approx70\,\mu$s \cite{PhysRevA.65.031401},
the strong gradient configuration yields
$\tau\approx87$ ms which is more than one order of magnitude
longer as the timescale of the envisaged dynamics.
Similarly, the van der Waals interaction of two Rydberg atoms
is suppressed by $|c|^4$.
Since the latter interaction is not accounted for in our calculations,
the resulting energy shift $\delta$ on the Rydberg levels should be 
marginal, i.e., well below the excitation detuning $\Delta_2$.
Taking $\delta<2\pi\times 1.0$ MHz as an example
(compared to $|\Delta_2|=2\pi\times50$ MHz)
yields a minimum interparticle distance of $\sim 250$ nm,
which is feasible even for condensate densities.

To conclude, we demonstrated that the composite nature of
Rydberg atoms is crucial for their trapping properties.
Our approach allows
to measure the specific features of the $nS_{1/2}$ Rydberg 
trapping potential by means of ground state atoms that are 
off-resonantly coupled to the Rydberg level.
Moreover, the proposed approach facilitates the designing of
trapping potentials for ground state atoms, which is demonstrated
by means of the example of a ring-shaped trap.

\begin{acknowledgments}
This work was supported by the German Research Foundation (DFG)
within the framework of the
Excellence Initiative through the Heidelberg Graduate School of
Fundamental Physics.
P.S.\ acknowledges fincancial support by the DFG, 
M.M.\ from the Landesgraduiertenf\"orderung
Baden-W\"urttemberg.
\end{acknowledgments}

\end{document}